\documentclass[prl,twocolumn,showpacs,preprintnumbers,amsmath,amssymb,euscript,amsfonts]{revtex4}
\usepackage{graphicx}% Include figure files
\usepackage{amssymb}
\usepackage{dcolumn}% Align table columns on decimal point
\usepackage{bm} % bold math
\usepackage[mathscr]{eucal}

\newcommand{\ie}{{\it i.e.}}

\def\ave#1{\langle#1\rangle}

\def\comm#1#2{[ #1,~#2]}

\begin{document}

\title{Unified Spin Order Theory via Gauge Landau-Lifshitz Equation}

\author{You-Quan Li, Ye-Hua Liu, Yi Zhou}
\affiliation{
% Zhejiang Institute of Modern Physics and
Department of Physics, Zhejiang University, Hangzhou 310027, P. R. China
}

\begin{abstract}

The continuum limit of the tilted SU(2) spin model is shown to
give rise to the gauge Landau-Lifshitz equation
which provides a unified description for various spin orders.
For a definite gauge, we find a double periodic solution,
where the conical spiral, in-plane spiral, helical, and
ferromagnetic spin orders become special cases, respectively.
For another gauge, we obtain the skyrmion-crystal solution.
By simulating the influence of magnetic field and temperature
for our covariant model,
we find a spontaneous formation of skyrmion-fragment lattice
and obtain a wider range of skyrmion-crystal phase in comparison to the conventional
Dzyaloshinsky-Moriya model.

\end{abstract}

%75.85.+t Magnetoelectric effects, multiferroics (for multiferroics and magnetoelectric films, see 77.55.Nv)
%75.10.Pq Spin chain models
%75.30.Gw Magnetic anisotropy
%03.65.-w Quantum mechanics
%75.10.Dg Crystal-field theory and spin Hamiltonians (see also 71.70.Ch Crystal and ligand fields)
%75.10.Hk Classical spin models
%75.25.Dk Orbital, charge, and other orders, including coupling of these orders
%75.30.Et Exchange and superexchange interactions (see also 71.70.Gm Exchange interactions)

\received{\today}

\maketitle

%--------Introduction--%

There has been spectacular progress in the study on the magnetoelectric
effects, which is expected to make a realistic step toward an electrical
control of magnetism\cite{Fiebig,Tokura06,Eerenstein,Choi}.
Within the intertwining of theory and experiment,
a mechanism based on spin-current showed that the
enhanced ferroelectric domains can be realized through cycloidal and
conical spin states in certain materials~\cite{Tokura06,Nag}.
For example, spiral spin state was shown~\cite{Nag} to cause electronic polarization.
Moreover, complex spin texture is interesting by its own right.
Recently, skyrmion lattice is observed in bulk MnSi~\cite{Muh}
and thin film $\mathrm{Fe}_x\mathrm{Co}_{1-x}\mathrm{Si}$~\cite{Tokura}.
Magnons in helical magnets are also probed by neutron
scattering experiments~\cite{heliband}.
Those nontrivial spin textures, undoubtedly,
play an important role in novel multiferroic materials.
There are mainly two kinds of mechanisms that cause spiral spin state,
one is ferro/antiferro-magnetic exchange competition
that is believed to be the origin of spiral state in manganites~\cite{j1j2j3},
the other is the antisymmetric Dzyaloshinsky-Moriya (DM) interaction
\cite{Dzyaloshinskii,Moriya} which arises from spin-orbit (SO) interaction
and manifests in crystal without inversion symmetry like MnSi.
Historically, Moriya~\cite{Moriya} was the first to give a microscopic treatment
of DM interaction based on Anderson's superexchange mechanism with SO interaction.
Thirty two years later, Shekhtman \cite{isodm} found that
Moriya's theory has a bond-isotropic form if only one takes all the
terms up to second order in SO interaction
which is usually neglected until present.
It is important to set up a unified description for
various spin orders.

In this letter, we indicate that
such a system can be described by tilted Heisenberg model
in which the tilting is related to the effects associated with bonds.
We formulate gauge Landau-Lifshitz equation from this model and
find solutions of various spin textures
and derive the dispersion relation of the relevant spin waves,
which provides a unified theory for spin orders
with insight in the gauge and geometric point of view.
Then we investigate the influence of external magnetic field and temperature
and plot the corresponding phase diagram
by making use of Monte Carlo simulations.

%------Introducing our model model--------%

In order to reach a unified description of various spin ordered phases
including the situations beyond the traditional ferromagnetic one,
we consider a much more generalized Heisenberg Hamiltonian
\begin{equation}
H = -J\sum_{\ave{jj'},c}U_j S^c_j U^{-1}_j \, U_{j'} S^c_{j'} U^{-1}_{j'}
\label{eq:tilt}
\end{equation}
where $c=1,2,3$,  $j\in L$ with $L$ the lattice space,
and $S^c_j$  denotes the $c$-th component of spin operator at site $j$.
These spin operators, proportional to the infinitesimal generators of SU(2),
obey
$\comm{S^a_j}{S^b_l} = i\hbar\delta_{jl}\epsilon^{abc}S^c_j$
that governs the time developments of any observable via Heisenberg equation of motion
for a definite model (\ref{eq:tilt}).
In Eq.~(\ref{eq:tilt}), $\ave{jj'}$ means the summation is taken over
the nearest neighbor lattice sites, and
the local tilting field $U_j$ accounts for
any effects arising from either (both) complicated crystalline fields or (and)
cumbersome charge order in whatever intricate materials.

As there exists a homomorphism between SU(2) and SO(3) Lie groups,
$U_j \boldsymbol{S}_j U^{-1}_j = \boldsymbol{S}_j O_j$
in which $\boldsymbol{S}_j$ denotes $(S^1_j, S^2_j, S^3_j)$
and $O_j$ the representation of SO(3),
each nearest-neighbor term in Eq.~(\ref{eq:tilt}) can be rearranged,
\ie,
$\displaystyle\sum_c U_j S^c_j U^{-1}_j \, U_{j'} S^c_{j'} U^{-1}_{j'}
=\boldsymbol{S}_j O_j(\boldsymbol{S}_{j'} O_{j'})^{\small T}
=\boldsymbol{S}_j O_j O_{j'}^{-1}\boldsymbol{S}_{j'}^{\small T}$.
Here $O_{j'}^{\small T}=O_{j'}^{-1}$ for orthogonal group has been used.
Because $j'$ is close to $j$ when the lattice constant $a$ is taken as an infinitesimal parameter,
we can expend $O_j O_{j'}^{-1}$
in the vicinity of identity,
namely
\begin{equation}
O_j O_{j'}^{-1}= 1 - a A^c_\nu (j)\,\hat{\ell}_c,
\label{eq:gauge}
\end{equation}
where we have considered the coordinate of site $j'$ is simply that of
$j$ plus  a bond vector $a\boldsymbol{e}_\nu$
in which $\boldsymbol{e}_\nu$ refers to the unit vectors
connecting neighborhood of a given lattice
structure.
Here
$\hat{\ell}_c$ denote the representation matrices of the infinitesimal generators of SO(3) Lie group,
they are $3\times 3$ matrices
$(\hat \ell_c)_{ab}= \epsilon_{abc}$
and fulfil the commutation relations~\cite{Gilmore}
$[\hat{\ell}_a\,, \hat{\ell}_b]=-\epsilon_{abc}\hat{\ell}_c$.
Clearly, the feature of the local tilting can be characterized by the SO(3)
non-Abelian gauge potential $\mathbb{A}_\nu (j) = A^c_\nu (j) \hat{\ell}_c$
which is a matrix valued vector field.
In order to avoid any ambiguity, here we clarify that
the $j$ represents a point in the lattice space
corresponding to the coordinate of real space in continuum model,
the $c$ labels the component of a vector in Lie algebra space while
the $\nu$ labels the one in real space.
Also for symbol neatness, in Eq.~(\ref{eq:gauge}) and thereafter,
we write the lattice-site label $j$ of
$\boldsymbol{A}$ in parentheses
rather than conventional subscripts.
By making use of Eq.~(\ref{eq:gauge}),
we can write Eq.~(\ref{eq:tilt}) as
\begin{align}
H = \frac{J}{2}\sum_{\ave{jj'}}
 \left[
   \Bigl(\boldsymbol{S}_{j'} - \boldsymbol{S}_{j}
         + a \boldsymbol{S}_j A^c_\nu (j)\,\hat{\ell}_c
   \Bigr)^2 - 2 C_j
 \right].
\label{eq:rewrite}
\end{align}
Actually, the Casimir invariants $C_j=\boldsymbol{S}_j\cdot\boldsymbol{S}_j=s_j (s_j +1)\hbar^2$
in a general system may differ at different lattice site,
which means the module of spin does not necessarily take the same value everywhere.
However, in this paper,
we focus on uniform spin module $S$ in every sites.

Now we are in the position to make continuum limit,
$\sum \rightarrow (1/a)^d \int d^d x $,
which can be realized by allowing the volume per lattice site $a^d$
tend to zero and
considering the lattice label $j$ as a continuous variable $\boldsymbol{r}$
and hence $\boldsymbol{S}_j$ as $\boldsymbol{M}(\boldsymbol{r})$.
Equation (\ref{eq:rewrite}) gives rise to  the effect Hamiltonian,
\begin{equation}
H=\frac{J}{2 a^{d-2}}\int d^d x
  \Big[ \bigl(\partial_\nu + \boldsymbol{A}_\nu(\boldsymbol{r}) \times \bigr)\boldsymbol{M}(\boldsymbol{r})\Bigr]^2,
\label{eq:hamiltonian}
\end{equation}
where the additional constant term is omitted.
Then the  corresponding Lagrangian density is given by
${\mathscr L}=a^{-d}|\boldsymbol{M}|(\cos\theta-1)\dot{\phi}
-a^{2-d}J/2\,(D\boldsymbol{M})^2 $
in which $(\theta, \phi)$ refer to the azimuthal angles of
$\boldsymbol{M}$.
The equation of motion for the spin field $\boldsymbol{M}(\boldsymbol{r}, t)$ is derived
as the following gauge Landau-Lifshitz equation,
\begin{eqnarray}
\frac{\partial}{\partial t} \boldsymbol{M}
 = a^2 J \boldsymbol{M}\times D^2\boldsymbol{M},
\label{eq:gauge-LLeq}
\end{eqnarray}
where $D^2=D_\nu D_\nu$ and the covariant derivative is given by
$D_\nu \boldsymbol{M}=(\partial_\nu + \boldsymbol{A}_\nu \times)\boldsymbol{M}$.
Equation (\ref{eq:gauge-LLeq}) is covariant under a gauge transformation
$\mathbb{A}_\nu \rightarrow G \mathbb{A}_\nu G^{-1} + \partial_\nu G G^{-1} $,
$(\boldsymbol{M})_a \rightarrow \sum_b G_{ab} (\boldsymbol{M})_b$
with $G \in \mathrm{SO(3)}$.

%-2D double periodic solution---

We first consider a typical gauge field in $x$-$y$ plane
$\boldsymbol{A}_x =(0, 0, -q_1)$,
$\boldsymbol{A}_y =(q_2\sin q_1 x, -q_2\cos q_1 x, 0)$.
We find a double periodic solution, $\boldsymbol{M}_\mathrm{dp}$,
as a steady solution of the gauge Landau-Lifshitz equation (\ref{eq:gauge-LLeq}),
\begin{eqnarray}
 \left\{
   \begin{array}{l}
m_1(x,y)=\sin(q_2 y +\beta)\cos(q_1 x ),
       \\[1mm]
m_2(x,y)=\sin(q_2 y +\beta)\sin(q_1 x ),
     \\[1mm]
m_3(x,y)=\cos(q_2 y +\beta).
    \end{array}
   \right.\label{eq:solution}
\end{eqnarray}
Here $\boldsymbol{m}=(m_1, m_2, m_3)$ refers to $\boldsymbol{M}_\mathrm{dp}/S$.
This spin order
is the exact ground-state solution of the system
because $D_{x}\boldsymbol{M}_\mathrm{dp}=D_{y}\boldsymbol{M}_\mathrm{dp}=0$
so that the positive definite energy functional (\ref{eq:hamiltonian})
reaches zero then.
Clearly, the conical spiral spin order~\cite{conic} is the special case of $q_2=0$
whose special case of $\beta=\pi/2$ reduces to the in-plane spiral spin order~\cite{inplane}.
The other cases $q_1=0$
or $q_1=q_2=0$ corresponds to a helical spin order~\cite{realhel}
or the ferromagnetic spin order, respectively.

To study the excitations above the aforementioned ground state
(\ref{eq:solution}),
we take  $\boldsymbol{M}_\mathrm{dp}+\delta{\boldsymbol M}$
and obtain the following linearized equation
$
\bigl(\partial_t - a^2 J\boldsymbol{M}_\mathrm{dp} \times D^2 \bigr)\delta{\boldsymbol M}=0
$.
Since the constraint $|\boldsymbol{M}|=S$ requires
$\boldsymbol{M}_\mathrm{dp}\cdot\delta\boldsymbol{M}=0$,
we can assume
$\delta\boldsymbol{M} = u(x, y, t)\boldsymbol{e}_\theta + v(x, y, t)\boldsymbol{e}_\phi$
with the local frame
$\boldsymbol{e}_\phi=\boldsymbol{e}_z \times \boldsymbol{M}_\mathrm{dp}/|\boldsymbol{e}_z \times \boldsymbol{M}_\mathrm{dp}|$,
$\boldsymbol{e}_\theta=\boldsymbol{e}_\phi \times \boldsymbol{M}_\mathrm{dp}
     /|\boldsymbol{e}_\phi \times \boldsymbol{M}_\mathrm{dp}|$.
Then the equations that possible low-lying excitation modes obey are
$\partial_t u + a^2 JS \nabla^2 v =0$ and
$\partial_t v - a^2 JS \nabla^2 u =0$.
Their Fourier transform gives rise to the dispersion relation
$\omega^2=a^4 J^2 S^2 |\boldsymbol{k}|^4.$
One can see that the dispersion relation here is happened to be the same
as that of the spin wave above a ferromagnetic ground state
in classical Heisenberg model.

Because the strength tensor
$F^c_{xy}= \partial_x A^c_y - \partial_y A^c_x + \epsilon^{abc}A^a_x A^b_y$
vanishes for the gauge potential relevant to the solution
(\ref{eq:solution}),
the gauge potential can be represented as a pure gauge
$\mathbb{A}_\nu = - G^{-1} \partial_\nu G$
with $G=\exp(q_2y\hat{\ell}_2)\exp(q_1x\hat{\ell}_3)$.
The generating matrix $G$ implies an important physical significance,
which transforms the double periodic spiral order (\ref{eq:solution})
to the traditional ferromagnetic order,
\ie,
$(\boldsymbol{M}_\mathrm{fe})_a = \sum_b G_{ab} (\boldsymbol{M}_\mathrm{dp})_b$.
Here $\boldsymbol{M}_\mathrm{fe}$ is the ground-state solution of Eq.~(\ref{eq:gauge-LLeq}) with null gauge potential.
The double periodic spiral order can be considered as a result of parallel displacement
of spin with the aforementioned gauge potential as connection.
Since the solutions referring to both orders are in the same equivalent class of gauge Landau-Lifishitz equation,
there would be no surprise that the dispersion relations for the excitations
above them are the same.

%---skyrmion lattice---%

Next, we investigate
the case with non-vanishing strength tensor,
which gives rise to skyrmion~\cite{Skyrme} crystal solutions.
For
$\boldsymbol{A}_x =(-\gamma/J, 0, 0 )$,
$\boldsymbol{A}_y =(0, -\gamma/J, 0 )$ where $\gamma$ denotes the strength of
spin-orbit interaction,
we have $\boldsymbol{F}_{xy}=(0, 0, \gamma^2/J^2)$
and energy density functional:
$
(J/2) \partial_\nu \boldsymbol{M}\cdot \partial_\nu \boldsymbol{M}
    +\gamma\boldsymbol{M}\cdot (\nabla\times\boldsymbol{M})
    +(\gamma^2/2J)\left[\boldsymbol{M}^{2}+(M_3 )^{2}\right]$.
Here the last term contributes an easy-plane anisotropy
that is the continuum version of Moriya's anisotropic exchange~\cite{Moriya},
the first two terms are the conventional ferromagnetic exchange and the
DM interaction which was used to explore possible states of skyrmion crystal~\cite{Tokura,Han}.
Unlike the solution of double periodic spiral order
which can be generated through a parallel displacement,
we need to solve the gauge Landau-Lifshitz equation at present.

For steady solution of the gauge Landau-Lifshitz equation (\ref{eq:gauge-LLeq}),
it is sufficient to solve the eigen-equation $D^2\boldsymbol{M} = \lambda\boldsymbol{M}$.
The $\lambda$ can be a scalar function in general
while it is assumed to be a constant here for simplicity.
It can be proven that the $\lambda$
is proportional to the energy density.
As being interested in periodic steady solution,
we can assume
$\boldsymbol{M}(\boldsymbol{r})=\boldsymbol{M}(\boldsymbol{k})\exp (i\boldsymbol{k}\cdot\boldsymbol{r})$.
Then the eigen-equations become a set of  algebraic equations
for us to determine $\boldsymbol{M}(\boldsymbol{k})$.
We obtain three solutions:
sinusoidal order with spin paralleling to the wave vector,
elliptically distorted right-handed and left-handed
helical order with spin perpendicular to the wave vector.
As the three eigenvalues depend on $|\boldsymbol{k}|^2$ merely,
we can make superposition of the eigenmodes corresponding to the same eigenvalue.
In the present case, $\gamma>0$ is assumed,
so we want chose the right-handed helical to construct the ground state.
The closest-packed lattice of skyrmions are superposition of
three such eigenmodes with three wave vectors
of the same length and mutually in $120^{\circ}$ angle, namely
$
\boldsymbol{M}_c=
\sum^3_{i=1} \boldsymbol{M}(\boldsymbol{k}_i) e^{i\boldsymbol{k}_i\cdot\boldsymbol{r}}
$
in which
$\boldsymbol{k}_{1}=(\frac{\sqrt{3}}{2}k,\frac{1}{2}k)$,
$\boldsymbol{k}_{2}=(-\frac{\sqrt{3}}{2}k,\frac{1}{2}k)$,
and
$\boldsymbol{k}_{3}=(0,-k)$.
Since the real and imaginary parts of $\boldsymbol{M}_{c}$ all satisfy Eq.~(\ref{eq:gauge-LLeq}),
we can normalize the real part to reach a physical state that is
a compromise of reducing energy and satisfying the unit-length constraint,
$
\boldsymbol{m}_\mathrm{sk}=\mathrm{Re}(\boldsymbol{M}_c)/|\mathrm{Re}(\boldsymbol{M}_c)|.
$
Here the unfixed parameter $k$ in $\boldsymbol{m}_\mathrm{sk}$
determines the lattice constant of the skyrmion crystal.

\begin{figure}[t]
\includegraphics[width=36mm]{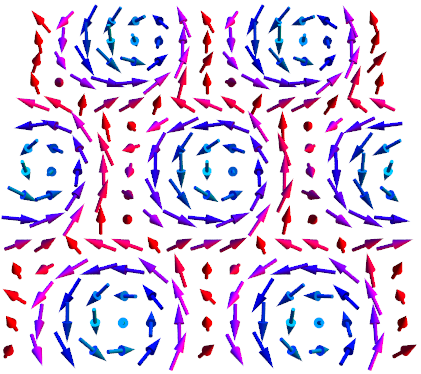}
\includegraphics[width=37mm]{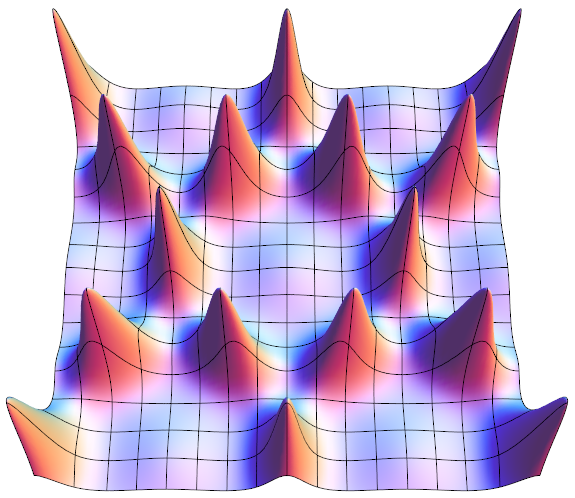}
\caption{(Color online)
Schematic illustration for spin order of skyrmion-crystal solution (left panel)
and the corresponding distribution of energy density in zero magnetic field (right panel).
The energy density is lower at the center of skyrmion
because the local spin chirality is large there that is favored by DM interaction;
it is larger in the boundaries of skyrmions
but the spins there favor a perpendicular magnetic field.}
\label{fig:skyrmion}
\end{figure}

The particular $k$ is determined by minimizing the average energy density
which is calculated through numerical integration.
Some features of the solution is plotted in Fig.~\ref{fig:skyrmion}
where spins between the center of skyrmions tend to
point up although the spins in each skyrmion tend to point down.
The average energy density of the optimized configuration of skyrmion crystal
is $0.276 S^2\gamma^2/J$ with $k=0.87\gamma/J$,
which is higher than helical order's $0.25 S^2\gamma^2/J$,
but the average $z$-component of spin for the solution $\boldsymbol{m}_\mathrm{sk}$ is $+0.17$.
The skyrmion crystal will have lower energy when
a sufficient large perpendicular magnetic field is applied downwards.
Whereas, when the magnetic field is further enhanced,
a ferromagnetic state with the $z$-component of spin
being $1$ eventually becomes the ground state.
This argument is consistent with Ref.~\cite{Tokura}.

%---anisotropy---%

Furthermore, we study what will happen if there exists a magnetic
anisotropy in the system.
Such an anisotropy can be introduced by adding the term $\sum_j \eta (S^z_j)^2$
in the spin model (\ref{eq:tilt}) in which either the easy axis is chosen as $z$-axis
or the easy plane as $x$-$y$ plane for  $\eta < 0$ or $\eta >0$, respectively.
Choosing this kind of anisotropy is due to
keeping the original rotational symmetry about the $z$-axis.
Then the above gauge Landau-Lifshitz equation
(\ref{eq:gauge-LLeq}) turns to the following anisotropic one,
\begin{eqnarray}
\frac{\partial}{\partial t} \boldsymbol{M}= a^2J \boldsymbol{M}\times D^2\boldsymbol{M}
  -2\eta \boldsymbol{M}\times\boldsymbol{M}',
\label{eq:gauge-aLLeq}
\end{eqnarray}
with $\boldsymbol{M}'=(0, 0, M_z)$.
Note that the gauge potential with skyrmion-crystal solution
merely contributes an easy plane anisotropy.
When the anisotropy coexist with the aforementioned gauge potential relevant
to the skyrmion-crystal solution,
the eigenequation for the original gauge Landau-Lifshitz equation in $k$-space is modified.
One can choose the right-handed helical mode
$\left(k_{y},-k_{x},i\rho\left|\boldsymbol{k}\right|\right)$
in which
$\rho=\sqrt{\xi^2+1}-\xi$, where $\xi=\left(\gamma^{2}+2J\eta/a^2\right)/
\left(4J\gamma\left|\boldsymbol{k}\right|\right)$.
In real space, this mode is a helical order of elliptic contour
with $\rho$ referring to ratio of semiminor and semimajor axes.
It can be seen that the larger the $\eta$ is, the smaller the $\rho$ will be,
which is in consistent with the requirement for minimizing the energy.
When $\xi=0$ we have $\rho=1$,
it occurs a cancelation between the added anisotropy term and
the second order term of $\gamma$ arising from the gauge potential.

%--finite temperature effect: a new order and phase diagram----%

\begin{figure}[t]
\includegraphics[width=29mm]{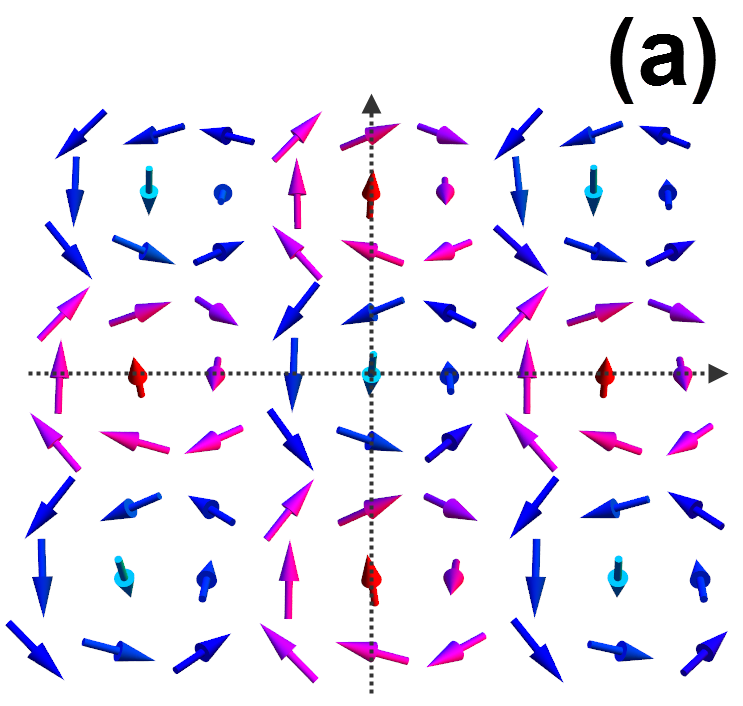}
\includegraphics[width=26mm]{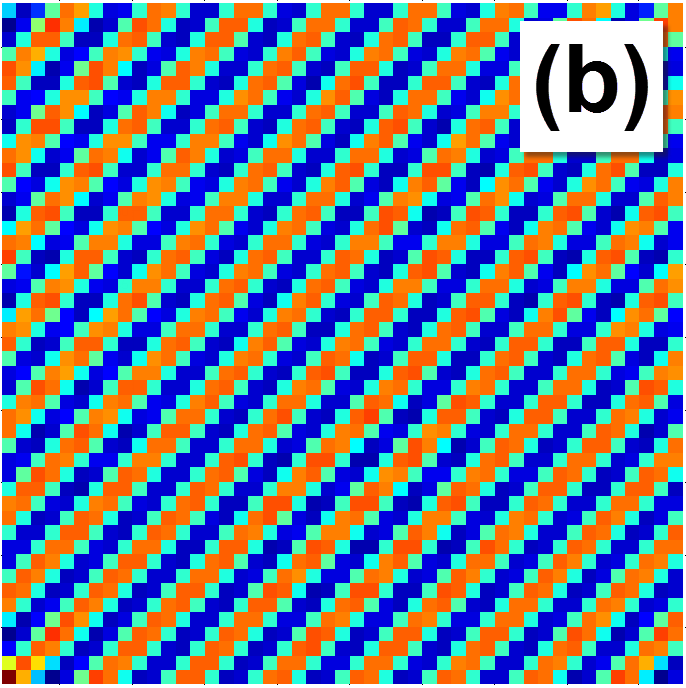}
\includegraphics[width=26mm]{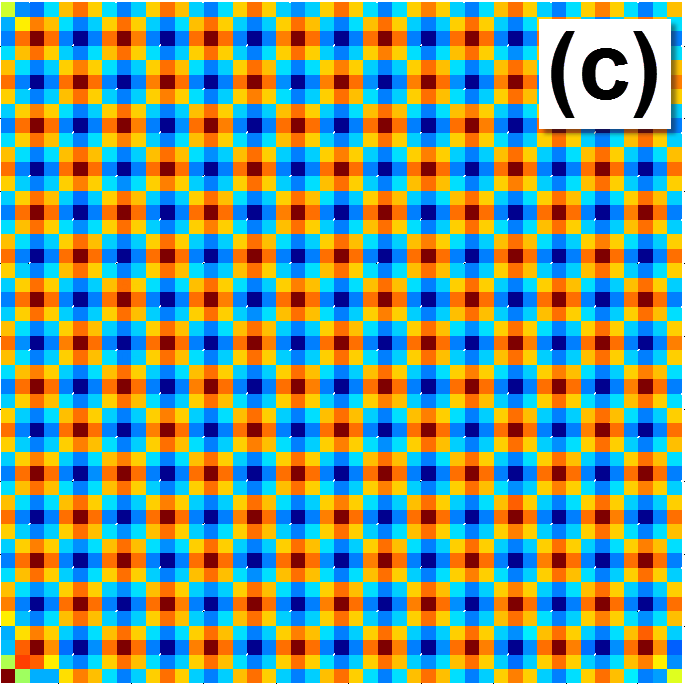}
\caption{(Color online)
({\bf a}) Spin order of skyrmion-fragment phase.
Here is a patch of spin textures from our simulation for $K/J=\sqrt{2}\tan{(2\pi/6)}$.
Along the black dot-line arrows the spins form a helical order
where each two successive spins have relative $60^\circ$ difference in angle.
({\bf b}) and ({\bf c}) The spin $z$-$z$ correlation in the 48 by 48 lattice
for various phases in zero magnetic field.
Temperature is lowering, it goes from disordered to helical (b) then to
skyrmion-fragment (c) phases.
}\label{fig:phase-spontaneous}
\end{figure}

\begin{figure}[t]
\includegraphics[width=83mm]{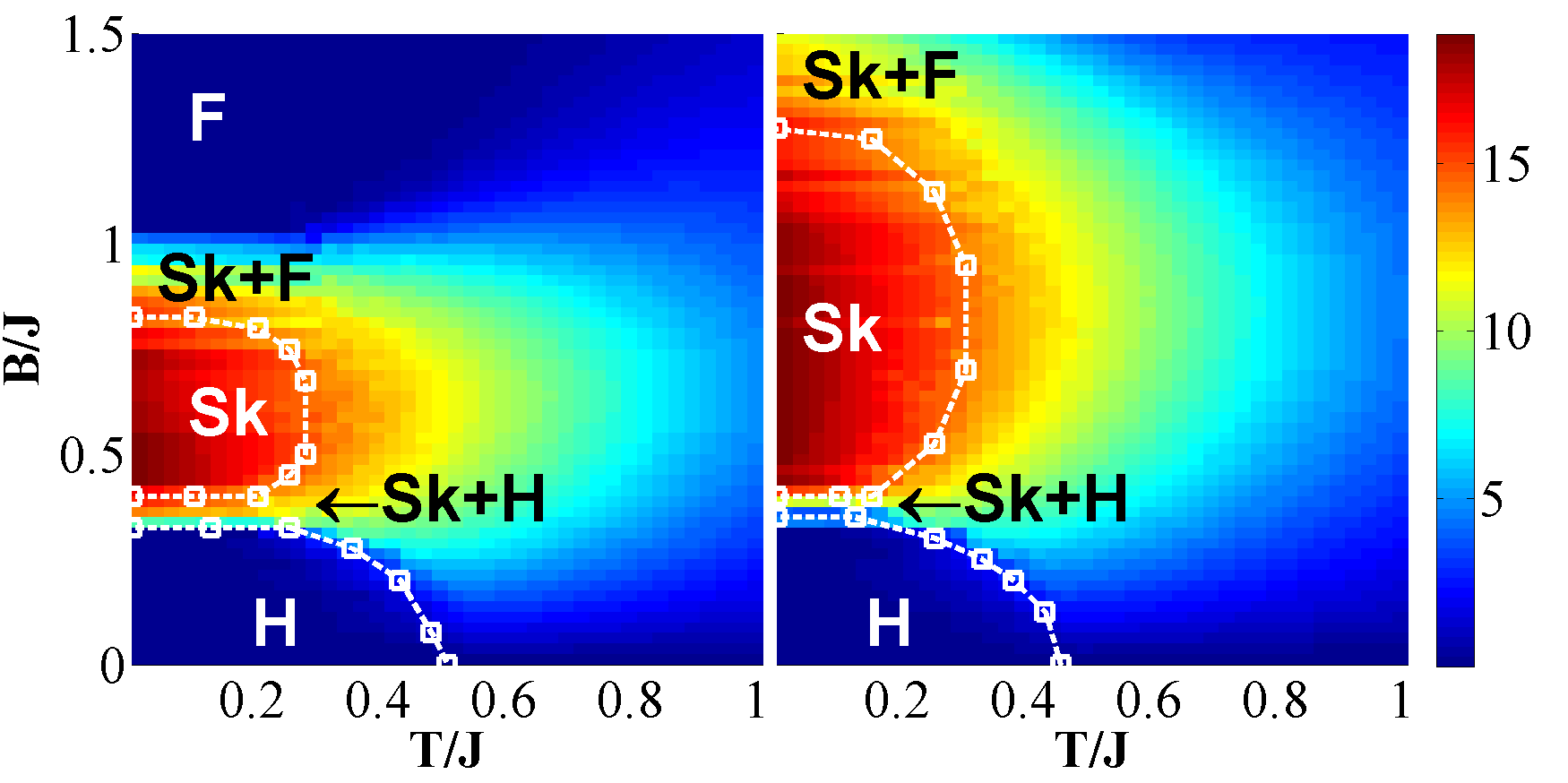}
\caption{(Color online)
Phase diagram of various spin orders
(Sk means skyrmion lattice, H helical, F ferromagnetic, and $+$ coexistence)
in the plane of magnetic field versus temperature 
calculated in the $36$ by $36$ lattice for DM model (left panel) and our covariant model (right panel).
The color indicates the total number of skyrmions. 
The helical phase and ferromagnetic phase have no skyrmions and the skyrmion lattice phase have many skyrmions. 
The area ratio of skyrmion phase to helical phase is larger in our covariant model.
}
\label{fig:phase-magnetic}
\end{figure}

Now we turn to investigate finite temperature effects
of our covariant model which contains DM interaction and magnetic anisotropy simultaneously.
For convenience in numerical simulation, we start from the lattice version,
\begin{align}\label{eq:simulation}
H = & \sum_{\boldsymbol{r},\boldsymbol{e}}
 \Bigl[-J\boldsymbol{S}_{\boldsymbol{r}}\cdot\boldsymbol{S}_{\boldsymbol{r}+a\boldsymbol{e}}
 -K\boldsymbol{e}\cdot\left(\boldsymbol{S}_{\boldsymbol{r}}\times\boldsymbol{S}_{\boldsymbol{r}+a\boldsymbol{e}
 }\right) \Bigr]
  \nonumber\\[0mm]
 & + \sum_{\boldsymbol{r}}\Bigl[
 \frac{K^{2}}{2J}\left(S_{\boldsymbol{r}}^{z}\right)^{2}-BS_{\boldsymbol{r}}^{z}\Bigr],
 \end{align}
where $\boldsymbol{S}$ denotes classical spin of unit module; 
$\boldsymbol{e}$ refers to $\hat{x}$ or $\hat{y}$
and $\boldsymbol{r}$ runs through the lattice site of the base space;
$J$, $K$ and $B$ denote the exchange, the strength of DM interaction and
the external magnetic field, respectively.
In our numerical calculation, 
the Boltzmann constant $k^{}_B$ and the lattice spacing $a$ is taken as unit.
We do Monte Carlo simulations in various regimes of model parameters.
For zero magnetic field $B=0$,
we find, for a specific strength of DM interaction $K/J=\sqrt{2}\tan{(2\pi/6)}$,
that the system goes from disordered phase to helical phase and then to a new phase
when temperature is lowering.
The new phase (see Fig.~\ref{fig:phase-spontaneous})
presents a square lattice of alternatively placed skyrmion fragments, some of which appear to be imbedded among spin helical textures that is marked by black dot-lines in Fig.~\ref{fig:phase-spontaneous}.
Since the new phase appears in zero magnetic field, it  
is an emergence of spontaneous formation of skyrmion-fragment lattice.
For weaker strength of DM interaction, saying $K/J=\sqrt{2}\tan{(2\pi/9)}$,
we plot the phase diagrams in the plane of temperature versus magnetic field
based on our Monte Carlo simulations for both our model (\ref{eq:simulation})
and the conventional model~\cite{Tokura}.
Our results manifest that
the landscape of those two phase diagrams  are similar
while the area ratio of skyrmion lattice phase to helical phase in our model is larger
than that in the conventional model (see Fig.~\ref{fig:phase-magnetic}).

%---Summary and 3D case ------%

In conclusion,
the gauge Landau-Lifshitz equation, as the continuum limit of the tilted SU(2) spin model,
provides a unified description for various spin orders.
The double periodic solution we found implies
the conical spiral, in-plane spiral, helical, and
ferromagnetic spin orders as special cases, respectively.
The skyrmion-crystal order is a solution corresponding to a
SO(3) gauge with nonvanishing strength tensor.
As to the finite temperature behavior,
a spontaneous formation of skyrmion-fragment lattice
occurs in zero magnetic field,
and the area ratio of skyrmion phase to helical phase is larger
in our covariant model than in the conventional DM model.
Note that the magnon band structure observed in recent experiments \cite{heliband}
does not contradict to double periodic dynamics
since it happens when the system is described by a
three-dimensional gauge potential
$\boldsymbol{A}_{x}=\left(-\gamma/J,0,0\right)$,
$\boldsymbol{A}_{y}=\left(0,-\gamma/J,0\right)$ and
$\boldsymbol{A}_{z}=\left(0,0,-\gamma/J\right)$.
This gauge potential is rotationally invariant and does not bring in anisotropy,
so our model is equivalent to the conventional model and
the ground state is the circularly helical state with wave vector $\gamma/J$.

We thank J.H. Han for useful communications.
The work is supported by NSFCs (grant No.11074216 \& No.11074218)
and  PCSIRT (Grant No. IRT0754).

\end{document}